\documentclass[conference]{IEEEtran}
\usepackage{amsmath, amssymb,multicol, graphicx,gensymb, flushend}

\IEEEoverridecommandlockouts

\begin{document}

\title{Plasma-Treated Polymeric Biomaterials for Improved Surface and Cell Adhesion}

\author{\IEEEauthorblockN{Jairo Rondón\IEEEauthorrefmark{1},
Angel Gonzalez-Lizardo\IEEEauthorrefmark{2}}
\IEEEauthorblockA{\IEEEauthorrefmark{1}Biomedical \& Chemical Engineering Departments,\\
Polytechnic University of Puerto Rico, San Juan, Puerto Rico, USA\\jrondon@pupr.edu}
\IEEEauthorblockA{\IEEEauthorrefmark{2}Department of Electrical and Computer Engineering and Computer Science,\\
Polytechnic University of Puerto Rico, San Juan, Puerto Rico, USA\\agonzalez@pupr.edu}
}

\maketitle

\begin{abstract}

Surface modification of polymeric biomaterials using plasma has emerged as an effective strategy to optimize the cell-material interface without compromising the structural properties of the material. This work presents a critical review of the impact of low-temperature plasma treatment on enhancing cell adhesion, with emphasis on the physicochemical changes induced on the surface of polymers commonly used in biomedical applications. The mechanisms of interaction between reactive plasma species and the polymer surface are analyzed, along with techniques used to introduce hydrophilic functional groups that improve wettability and biocompatibility. Scientific evidence demonstrates that this type of surface modification promotes greater cell spreading, anchorage, and proliferation, making it particularly useful in the design of tissue engineering scaffolds, implantable devices, and vascular prostheses. Finally, current and future trends in the development of smart plasma-functionalized biomaterials are discussed, highlighting their role in regenerative medicine.
\end{abstract}

\section{Introduction}

The interaction between biomaterials and biological tissues represents a fundamental axis in the design and development of medical devices, tissue engineering scaffolds, and both permanent and temporary implants. In particular, polymeric biomaterials have gained a prominent role in this field due to their versatility, biocompatibility, ease of processing, and tunable mechanical properties \cite{r1}, \cite{r2}. However, despite these advantages, a major limitation of these materials lies in their surface properties, which often lack the necessary bioactivity to induce optimal cellular responses such as adhesion, proliferation, and differentiation \cite{r3}.

Given that the surface of a biomaterial is the first interface to come into contact with the biological environment, its physicochemical characteristics---such as chemical composition, surface energy, roughness, and wettability---play a critical role in modulating cellular behavior \cite{r4}, \cite{r5}. Consequently, surface modification strategies have emerged as a promising approach to enhance the biological functionality of biomaterials without altering their internal structural properties \cite{r6}.

Among these strategies, plasma surface modification (PSM) has positioned itself as an efficient, clean, and environmentally friendly technology for selectively altering the surface of polymeric materials. This technique allows for the incorporation of active functional groups---such as hydroxyl, carboxyl, and carbonyl groups---that significantly enhance cell-material interactions, all without the need for harsh chemical reagents or subsequent purification steps \cite{r1}, \cite{r5}. Furthermore, plasma treatment enables precise control over surface chemistry, surface free energy, and topographical features, thereby promoting improved biological integration and specific cellular responses \cite{r7}, \cite{r2}.

In this context, the present article aims to provide a critical review of recent advances in the use of plasma surface modification in polymeric biomaterials, with a particular focus on enhancing cell adhesion---a key parameter for success in tissue engineering and regenerative medicine applications.

\section{Methodology}

The methodology applied in this research involved conducting a critical and comprehensive review of scientific literature to collect information related to tissue engineering principles, surface modification techniques for biomaterials, recent advances in plasma treatment, and current challenges in biomedical material design.

\begin{enumerate}
    \item Search and data compilation:
A broad search was carried out using reputable scientific databases, including PubMed, Frontiers, UPComing, Wiley Online Library, Royal Society of Chemistry, MDPI, ACS Publications, ScienceDirect, SCOPUS, IEEE, SciELO, IOP, RedALyC, and Google Scholar. Key search terms included “tissue engineering”, “biomedical engineering”, “plasma surface modification”, “polymeric biomaterials”, and “cell adhesion”. The search covered publications from 1993 to 2024, ensuring both foundational and contemporary works were considered.
    
    \item Information selection and refinement:
After the initial data collection, sources were refined based on scientific relevance and direct applicability to the subject matter. The Mendeley reference manager \cite{r8} was used to organize, classify, and cite the literature. Priority was given to articles focusing on plasma modification, cell-material interactions, and biomedical applications of polymeric materials.
    
    \item Thematic structuring and subtopic selection: 
    The refined information facilitated the logical structuring of the research and helped identify key subtopics, including: fundamentals of plasma surface modification, limitations of conventional biomedical polymers, physicochemical mechanisms induced by plasma treatment, effects on cell adhesion, and emerging clinical applications.    
    \item Critical analysis and conclusion development: A thorough and integrative analysis of the selected literature was conducted to identify trends, compare findings, and formulate evidence-based conclusions on the efficacy of plasma treatment on polymeric biomaterials. These insights are discussed in detail throughout the study \cite{r9}.

\end{enumerate}

\section{Discussion and Results}

\subsection{Fundamentals of Plasma Surface Modification}
Plasma surface modification (PSM) has emerged as an efficient, versatile, and environmentally sustainable technology for tailoring the surface properties of biomaterials without compromising their bulk structure. Plasma, defined as a partially ionized gas composed of electrons, ions, free radicals, and excited species, has the ability to interact with polymer surfaces and induce physicochemical transformations at the nanometric scale \cite{r4}, \cite{r5}. These modifications are essential for optimizing biocompatibility and promoting specific interactions with cellular components.

Various plasma configurations are employed in the modification of biomaterials, among which low-pressure plasma systems (LPPS) and atmospheric pressure plasma systems (APPS) are particularly prominent. Both approaches enable surface activation through the incorporation of polar functional groups such as hydroxyl (–OH), carbonyl (–C=O), and carboxyl (–COOH), as well as by increasing surface energy—factors that enhance protein adsorption and cell adhesion \cite{r1}, \cite{r2}, \cite{r7}.

The PSM process involves multiple mechanisms, including surface cleaning, chemical activation, physical etching, and the deposition of thin functional layers. During the activation phase, reactive plasma species break chemical bonds on the polymer surface, generating radical sites that allow for subsequent functionalization or the grafting of new chemical structures \cite{r4}, \cite{r10}. These changes can significantly alter the wettability of the material—a parameter closely linked to cell adhesion. It has been documented that a water contact angle close to 70° provides optimal conditions for cell-material interactions \cite{r1}, \cite{r11}, \cite{r12}.

A key advantage of plasma treatment is its non-invasive nature and avoidance of organic solvents, which minimizes toxicological risks and prevents the formation of harmful byproducts, making it especially suitable for sensitive biomedical applications. Moreover, this technique allows precise control over treatment variables such as gas type, pressure, power, and exposure time, facilitating the design of functional surfaces tailored to specific clinical needs \cite{r5}, \cite{r13}, \cite{r14}.

\subsection{Polymeric Materials and Surface Limitations}

Synthetic polymers are widely used in biomedical applications due to their versatility, ease of processing, tunable mechanical properties, and relative biocompatibility. Commonly employed materials include polyethylene (PE), polypropylene (PP), polyethylene terephthalate (PET), polytetrafluoroethylene (PTFE), polyurethane (PU), polycaprolactone (PCL), polylactic acid (PLA), and copolymers such as PLGA (polylactic-co-glycolic acid). These polymers have demonstrated reliable performance in various applications, including tissue engineering scaffolds, implantable devices, prosthetics, and controlled drug delivery systems \cite{r2}, \cite{r5}.

Despite their advantages, many of these polymers exhibit hydrophobic, chemically inert surfaces with low surface energy, which limits protein adsorption as well as subsequent cell adhesion and proliferation \cite{r1}. These interfacial deficiencies can compromise the clinical performance of the material, leading to poor integration with surrounding tissue and increasing the risk of fibrous encapsulation or implant rejection.

For instance, low-density polyethylene (LDPE) and PTFE exhibit water contact angles greater than 90°, indicating highly hydrophobic surfaces that are unfavorable for cell adhesion. Even moderately hydrophobic polymers such as PET and PU can benefit from surface modification strategies to improve their bioactivity \cite{r4}. Studies have shown that the incorporation of polar functional groups via plasma treatment significantly enhances the ability of these polymers to interact with extracellular matrix proteins such as fibronectin and collagen, which are critical for cell anchorage and expansion \cite{r1}, \cite{r5}.

In addition, some polymers present other surface-related limitations, such as unfavorable topography, static surface charge, or the tendency to generate undesirable degradation products. These characteristics can interfere with the formation of appropriate cellular microenvironments, hindering cell migration, differentiation, and functional behavior \cite{r15}.

In this context, plasma surface modification has emerged as an effective solution for overcoming these challenges, enabling the specific functionalization of the polymer interface without altering its mechanical properties or internal structure. This customization capability has been essential in adapting polymeric materials to the specific demands of regenerative medicine and tissue engineering \cite{r13}, \cite{r14}.

\subsection{Plasma-Induced Changes on Polymeric Surfaces}

Plasma surface treatment enables both chemical and physical modifications of polymer surfaces without compromising their internal structural properties. These alterations result from the interaction between reactive plasma species—such as ions, free radicals, electrons, photons, and excited molecules—and the atoms located in the outermost layer of the material \cite{r2}, \cite{r4}.

From a chemical standpoint, one of the main effects of plasma treatment is the incorporation of polar functional groups such as hydroxyl (-OH), carbonyl (-C=O), carboxyl (-COOH), and amine (-NH$_2$) groups. These moieties increase the surface energy of the polymer, enhance wettability, and facilitate the adsorption of extracellular proteins \cite{r1}, \cite{r5}. This surface functionalization promotes the formation of specific bonds between cell adhesion proteins and membrane receptors, such as integrins, thereby enabling more efficient cellular adhesion.

From a physical perspective, plasma can also alter surface roughness through controlled etching or ablation processes. An increase in nano- and micro-scale roughness provides more anchoring points and greater contact area for cells, improving adhesion \cite{r4}. This topographical enhancement has been linked to better cytoskeletal organization, formation of focal adhesions, and increased cell proliferation.

Additionally, plasma exposure can generate surface-bound free radicals, enabling the subsequent covalent immobilization of biomolecules such as proteins, bioactive peptides, or growth factors—further enhancing the bioactivity of the treated material \cite{r5}.

It is important to note that the effects of plasma treatment depend on several variables: the type of gas used (e.g., oxygen, nitrogen, argon, air, or reactive gas mixtures), plasma power, system pressure, exposure time, and the chemical nature of the polymer \cite{r21}. For example, oxygen plasmas are effective in introducing carboxylic groups, while ammonia or nitrogen plasmas are more suitable for introducing amine functionalities \cite{r23}.

These plasma-induced changes are generally confined to the outermost layers of the material (1--100~nm), which helps preserve the mechanical and thermal properties of the biomaterial. However, some of these effects may be transient due to surface rearrangement or post-treatment environmental contamination. Therefore, post-treatment stabilization or additional functionalization is sometimes required \cite{r24}.

Hence, plasma surface modification is a powerful tool for optimizing the cell-material interface, thanks to its ability to precisely tailor the chemistry, energetics, and surface topography of polymers.

\subsection{Effects of Plasma Treatment on Cell Adhesion}

Cell adhesion is a fundamental process in the interaction between biomedical materials and biological tissues, and its effectiveness is largely determined by the physicochemical properties of the biomaterial surface. Plasma treatment has proven to be an effective technique for significantly enhancing cell adhesion on polymeric substrates by inducing changes that create a more favorable microenvironment for cellular anchorage \cite{r1}, \cite{r5}.

First, the incorporation of polar functional groups such as hydroxyl, carbonyl, and carboxyl groups increases the surface energy and improves the wettability of the material. These changes facilitate the adsorption of extracellular matrix (ECM) proteins—such as fibronectin, vitronectin, and collagen—which are essential for mediating the initial cell-surface interaction \cite{r4}. The favorable adsorption of these proteins allows for the specific binding of cellular integrins, promoting the formation of stable focal adhesions and cytoskeletal reorganization.

Second, plasma-induced modification of surface topography, particularly at the nanoscale level, increases the effective cell contact area and provides additional anchorage sites for cellular filopodia. Studies have shown that surfaces with controlled roughness generated by plasma promote greater cell spreading, lamellipodia formation, and differentiation—particularly in osteoblastic, endothelial, and fibroblastic cells \cite{r2}.

Moreover, the creation of active radical-rich surfaces via ion bombardment facilitates the covalent immobilization of bioactive molecules that modulate cell adhesion and proliferation, such as RGD peptides, growth factors, or cytokines. These functionalized surfaces enhance cellular responses in specific applications, including tissue regeneration and implantable device design \cite{r5}.

A key parameter identified by Tamada \cite{r1} is the optimal contact angle for cell adhesion, which is approximately 70\textdegree. This value represents a balance between sufficient hydrophilicity to support orderly protein adsorption and enough hydrophobicity to prevent protein denaturation. A well-controlled plasma treatment can precisely adjust this contact angle, thereby optimizing the material's biointeraction.

It is worth noting that the beneficial effects of plasma treatment on cell adhesion have been observed in a variety of biomedical polymers, including polyethylene (PE), polytetrafluoroethylene (PTFE), polyethylene terephthalate (PET), polystyrene (PS), and polypropylene (PP). However, the magnitude of the effect depends on the chemical nature of the polymer, the type of plasma used, and the specific treatment conditions.

Therefore, plasma surface modification improves cell adhesion through a synergistic combination of chemical, topographic, and energetic changes at the polymer surface. This phenomenon is crucial for the design of advanced biomaterials capable of effectively integrating into the biological environment.

\section{Biomedical Applications of Plasma-Treated Polymeric Biomaterials}

Plasma surface modification has significantly expanded the range of biomedical applications for polymeric biomaterials, owing to its ability to enhance cell-material interactions without compromising the mechanical and structural properties of the substrate. Due to its versatility, controllability, and environmentally friendly nature, plasma treatment has become a strategic tool for the development of advanced medical devices, functional implants, and scaffolds for tissue engineering \cite{r2}, \cite{r5}.

One of the main applications is in the design of scaffolds for tissue engineering, where cell adhesion, proliferation, and differentiation are essential for functional tissue regeneration. For instance, plasma modification of polymers such as polylactic acid (PLA), low-density polyethylene (LDPE), and poly($\epsilon$-caprolactone) (PCL) has been shown to significantly enhance cell colonization and extracellular matrix (ECM) formation-critical factors for applications in bone, cartilage, and vascular tissues \cite{r12}, \cite{r16}, \cite{r17}.

In vascular grafts and cardiovascular devices, plasma treatment improves hemocompatibility by reducing nonspecific plasma protein adsorption and minimizing platelet activation. Furthermore, it enables the immobilization of anticoagulants such as heparin or biomolecules that promote endothelialization, resulting in blood-compatible surfaces and reduced thrombogenic risk \cite{r6}, \cite{r18}.

Implantable devices, including catheters, intraocular lenses, and joint prostheses, also benefit from plasma treatment, which reduces bacterial adhesion and enhances tissue integration. For example, surface modification of polymethyl methacrylate (PMMA) and polycarbonate (PC) has been reported to decrease biofilm formation and promote favorable immune responses \cite{r12}, \cite{r19}, \cite{r20}.

An emerging application lies in controlled drug delivery systems, where plasma treatment allows for the selective modification of polymer surfaces without altering their internal structure. This technique has been employed to generate functionalization gradients that regulate the adsorption and release of therapeutic agents, proving especially useful in dermal patches, polymeric microcapsules, and bioactive coatings \cite{r13}, \cite{r14}.

Finally, in the field of biosensors and diagnostic systems, plasma enables the precise functionalization of polymer surfaces for the immobilization of antibodies, enzymes, or nucleic acids without the need for harsh chemical reagents. This capability has facilitated the development of more sensitive, selective, and stable devices, with applications in clinical monitoring, early diagnosis, and personalized medicine \cite{r22}.

Collectively, these applications highlight the high potential of plasma treatment to enhance the performance and biocompatibility of polymeric biomaterials, offering innovative solutions for current challenges in regenerative medicine, implantology, and the design of intelligent medical devices.

\section{Conclusion}

Plasma surface modification has established itself as a powerful and versatile technique for optimizing the surface properties of polymeric biomaterials without compromising their structural integrity or mechanical performance. By introducing functional groups such as hydroxyl, carbonyl, and carboxyl moieties, it enables significant improvements in surface energy, wettability, and biocompatibility—key factors in promoting cell adhesion, proliferation, and differentiation.

Findings reported in the scientific literature demonstrate that plasma treatment can transform inert polymers such as polyethylene, polypropylene, and polystyrene into bioactive surfaces suitable for advanced biomedical applications. Moreover, this technique facilitates the development of smart devices and scaffolds capable of actively interacting with the biological environment, thereby enhancing tissue integration, the functionality of implantable devices, and the efficacy of controlled drug delivery systems.

Finally, plasma surface modification represents a strategic tool for the design of next-generation polymeric biomaterials, with promising applications in tissue engineering, regenerative medicine, implantology, cardiovascular devices, and biosensors. Its implementation in biomedical practice may lead to more effective, safer, and personalized solutions in alignment with the emerging trends of future medicine.

\end{document}